\documentclass[prb, twocolumn, superscriptaddress]{revtex4-2}
\usepackage{amsmath,amssymb,bm}
\usepackage{hyperref}
\usepackage{graphicx}
\usepackage{epstopdf}
\usepackage{latexsym}
\usepackage[usenames, dvipsnames]{color}
\usepackage[usenames, dvipsnames]{xcolor}
\usepackage{braket}
\usepackage{float}
\usepackage[normalem]{ulem}
\usepackage{comment}
\usepackage{mathtools}
\usepackage{array}
\usepackage{tabu}
\usepackage{multirow}
\usepackage[inline]{enumitem}
\usepackage{cleveref}
\usepackage{xcolor}
\usepackage[normalem]{ulem}
\usepackage{natbib}
\usepackage{amsmath}

\begin{document}

\title{Spontaneous fractional Josephson current  from parafermions}
\author{Kishore Iyer}
\affiliation{Aix Marseille Univ, Université de Toulon, CNRS, CPT, Marseille, France}
\affiliation{International Centre for Theoretical Sciences, Tata Institute
of Fundamental Research, Bengaluru 560089, India}
\author{Amulya Ratnakar}
\affiliation{Department of Physics, Indian Institute of Science Education and Research (IISER) Kolkata, Mohanpur - 741246, West Bengal, India}
\email{ORCID ID: 0000-0002-0025-9552}
\author {Aabir Mukhopadyay}
\affiliation{Department of Physics, Indian Institute of Science Education and Research (IISER) Kolkata, Mohanpur - 741246, West Bengal, India}
\email{ORCID ID: 0000-0001-6465-2727}
\author{Sumathi Rao}
\affiliation{International Centre for Theoretical Sciences, Tata Institute
of Fundamental Research, Bengaluru 560089, India}
\author{Sourin Das}
\affiliation{Department of Physics, Indian Institute of Science Education and Research (IISER) Kolkata, Mohanpur - 741246, West Bengal, India}

\setcounter{affil}{0}
\renewcommand{\thefigure}{\arabic{figure}}
\setcounter{figure}{0}
\renewcommand{\theequation}{\arabic{equation}}
\setcounter{equation}{0}
\renewcommand\thesection{\Roman{section}}
\setcounter{section}{0}

\begin{abstract}
We study a parafermion Josephson junction (JJ) comprising a pair of counter-propagating edge modes of two quantum Hall (QH) systems, proximitized by an s-wave superconductor. We show that the difference between the lengths (which can be controlled by external gates) of the two counter propagating chiral edges at the Josephson junction, can act as a source of spontaneous phase bias. For the Laughlin filling fractions, $\nu = 1/m,~ m \in 2\mathbb{Z}+1$, this leads to an electrical control of either Majorana $(m=1)$ or parafermion $(m\neq 1)$ zero modes. 
\end{abstract}

\maketitle

Parafermions~\cite{Fendley2012_parafermions,Fidkowski2010,AliceaReview_parafermions,Lee2017_parafermion,Vaezi13_parafermions,Lindner12_parafermion,Clarke2013_parafermions,udit&yuval_parafermion}  are   exotic  generalizations of the  Majorana modes~\cite{kitaev2001_majorana, read2000_majorana, nilsson2008_majorana, fu2009_majorana, oreg2010_majorana, das2012_majorana, lutchyn2010_majorana, alicea2012_majoarana, sarma2015_majorana, bommer2019_majorana} which may give rise to topological qudits - higher dimensional generalizations of qubits -  with an even better fault tolerance\cite{kitaev2001_majorana, Cheng_majorana, preskill2012}  than Majorana qubits.  The essential property of these exotic excitations that make them relevant for topological quantum computation is their behavior under exchange -- they transform as non-abelian anyons. Non-abelian anyons are higher dimensional representations of the braid group where exchanges are represented by unitary matrices, which do not commute. So exchanging parafermions or braiding them will essentially rotate them in the Hilbert space of the degenerate ground state manifold. This nonlocal nature of operations generated by non-abelian braiding gives rise to fault-tolerance, making systems hosting non-abelian anyons promising  platforms for quantum information processing.

Majorana modes, also called Ising anyons, are the simplest examples of excitations that have  non-abelian braiding statistics. This has spearheaded the experimental search for these  Majorana modes, which  have now been expanded to many different platforms such as one-dimensional wires~\cite{Sach_wire,cook2011_wire,Halperin_wire, oreg2014_wire,clarke2011_wire,stanescu2013_wire, fidkowski2011_wire,kjaergaard2012_wire, prada2020_wire, transport_setiawan,conductance_setiawan,Topologica_SC_setiawan}, fractional Josephson effect experiments\cite{Clarke2013_parafermions,cheng2015_FJJ,Cheng12,Jian_FJJ,CCkai_FJJ,frolov2020_FJJ,rokhinson2012_FJJ, alidoust1, alidoust2}, etc.
There is a growing consensus in the community  that there exists incontrovertible experimental evidence for Majoranas, despite some drawbacks of the evidence \cite{castelvecchi2021}.

Experimental searches  for parafermion detection, on the other hand, are still in their infancy. 
Even the minimal proposals for the detection of parafermions involve a pair of FQH edge states or edge states of a fractional topological insulator - $i.e.$, even the simplest proposals involve strong electron-electron interactions. By now, there
exist  several proposals to engineer parafermions involving multiple or multi-layer FQH states or fractional topological insulator states proximitized by superconductors (and/or ferromagnets)\cite{Vaezi13_parafermions, Lindner12_parafermion, Clarke2013_parafermions, Cheng12, Mong14,Barkeshli14PRX,oreg_parafermions}. There has even been  experimental evidence\cite{Lee2017_parafermion,gul2021andreev_parafermion} of crossed Andreev reflection of fractionally charged edge states in a graphene based FQH system proximitized with a superconducting lead, and more recently in semiconductor IQH systems \cite{shabani22}, which are precursors to being able to localize parafermions.

In this letter, our main focus is to re-examine the fractional Josephson effect that occurs when the edges of a quantum spin Hall insulator or FQH states are sandwiched between two superconductors, but  with one important difference. 
We allow for the two edges to have independent gate-tunable lengths $L_1$ and $L_2$. Even for a quantum spin Hall system, where the edge states can be described by free electrons, and the spectrum of the Andreev bound states shows $4\pi$
fractional Josephson effect, 
we find that the finite independent lengths have important consequences and lead to a spontaneous Josephson current even in the absence of a phase difference. 
These consequences persist when the state between the superconductors are two independent $\nu=1/m$ fractional quantum Hall states, and we obtain an
appropriate spontaneous fractional Josephson current as a function of the difference in the lengths of the two edges.

\noindent\emph{ The Majorana case:} \label{majorana}

\begin{figure}
\centering
\includegraphics[scale=0.13]{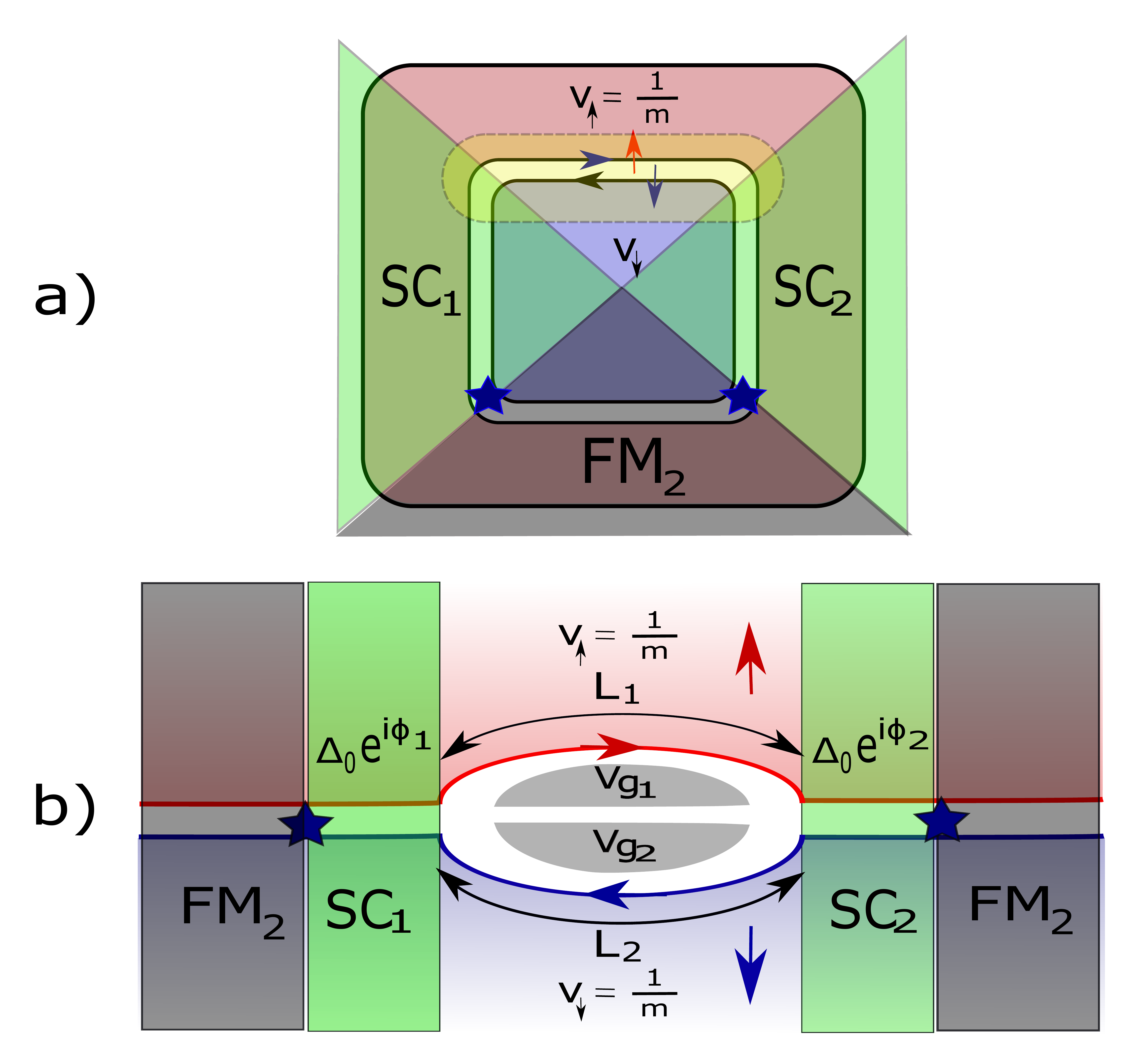}
\caption{Caricature of an idealized experimental set-up. Fig. 1(a) shows two concentric FQH liquids at filling fractions $\nu_{\uparrow/\downarrow}=1/m$, ($m \in$  odd integer), colored red/blue respectively, with counter-propagating edge modes and opposite spins. The edge modes are proximitized by two superconductors, $SC_{1}$ and $SC_{2}$, colored green, and a ferromagnet ${FM_{2}}$ colored grey. The encircled (yellow) region comprises the free edges and the magnified version of this is shown in Fig.1(b). $V_{g_{1/2}}$ are gate potentials which can individually alter the length of the edges in the free region. $L_{1/2}$ are the lengths of the right moving and left moving edge modes, respectively. $\Delta_{0}$ and $\phi_{i}$ are the superconducting gaps and the superconducting phases corresponding to $SC_i$. The two superconducting segments are considered to be the part of the same bulk superconductor. The blue stars at the interface between $SC_{i}$ and $FM_{2}$ represent localized parafermion zero modes.} 
\label{set-up}
\end{figure}

The junction between the two QH edge states described in fig.~\ref{set-up} allows for realization of a helical edge state \cite{sanchez2017_HelEdge,ronen2018_HelEdge}, which
when proximitized by the superconductors leads to a topological phase with effective p-wave
superconducting correlations~\cite{alicea2012_majoarana}. The ballistic Josephson junction hence formed is expected to show $4\pi$ periodic Josephson effect, provided that fermion parity is preserved~\cite{fu2009_majorana}. Further, we will allow the counter-propagating edges in the ballistic region to have different lengths ($L_1$ and $L_2$), which may be realized by appropriate gating, as shown schematically in fig.~(\ref{set-up}b).

Since, for $m=1$, the quasi particles at the edge
are essentially free electrons, we can write the Hamiltonian for the quantum  Hall edges proximitized by superconductors and ferromagnets as $H = H_0 + H_I$ where
\begin{eqnarray}
H_0 &=& -i\hbar v_F\int dx~\Big[ \psi_R^\dagger(x) \partial_x \psi_R(x) + \psi_R(x)\partial_x\psi_R^\dagger(x)\Big] \nonumber\\   
&& + i\hbar v_F \int dx~ \Big[ \psi_L^\dagger(x)\partial_x\psi_L(x) + \psi_L(x)\partial_x\psi_L^\dagger(x) \Big]
\nonumber\\ 
H_{I} &=& \int dx \left( \Delta(x)\psi_{R}\psi_{L} + M(x)\psi^{\dagger}_{R}\psi_{L} +h.c. \right)
\end{eqnarray}
where $\psi_{R/L}$ are right/left-moving chiral fermionic fields and $v_F$ is the Fermi velocity of the electrons in these edges. The pairing amplitude $\Delta(x)$ and the backscattering strength $M(x)$ have the spatial profile, determined by the set-up. The presence of superconducting correlations on a finite patch of the fermionic edges can be reduced to Andreev boundary conditions on the edges of the fermionic fields in the free region of the set-up~\cite{Maslov96,stone2011_ABS,crepin2014_ABS,crepin_trauzettel,tanaka1999_ABS,mukhopadhyay_ABS,Kundu2010} as shown below -
\begin{eqnarray}
\psi_{R,\uparrow}(x=0) = e^{-i\Phi}e^{i\phi_{1}}\psi^{\dagger}_{L,\downarrow}(x=0) \nonumber\\
\psi_{R,\uparrow}(x=L_{1}) = e^{-i\Phi}e^{i\phi_{2}}\psi^{\dagger}_{L,\downarrow}(x=L_{2})
\label{Eq: psi_BC}
\end{eqnarray}
where $\Phi = \text{cos}^{-1}\left( \frac{E}{\Delta_0} \right)$, $E$ is the ABS energy, {and $\phi_1$ and $\phi_2$ are the phases of the two superconducting regions}.  The boundary condition assumes that the superconductors are wide enough so that the Majorana modes  localized at the interface between $SC_{1/2}$ and $FM_2$ do not influence it.
The ABS spectrum can then be easily calculated to be \cite{Kundu2010} (see Supplemental material) 
\begin{equation}
    E = \pm \Delta_{0} \text{cos}\Bigg[ \frac{E}{\Delta_{0}}
    \frac{\langle L \rangle}{L_{SC}} \pm \Big(\frac{\mu\delta L}{\hbar v_F} -\frac{\phi}{2}  \Big) \Bigg] \label{andreevspectrum}
\end{equation}
where $\mu$ denotes the Fermi energy, $\langle L \rangle = \frac{L_1 + L_2}{2}$, $\delta L = \frac{L_1 - L_2}{2}$, $\phi = \phi_1 - \phi_2$ {is the difference of the two superconducting phases} and $L_{SC} = \hbar v_{F}/\Delta_{0}$ is the superconducting coherence length. In the short junction limit, that is, $ L_{1/2}/L_{SC} \longrightarrow 0$, Eq.~\ref{andreevspectrum} reduces to the well known ABS energy for a ballistic junction, given by, $E = \pm \Delta_{0}\cos{\phi/2}$~\cite{fu2009_majorana,mukhopadhyay_ABS,Kundu2010, kwon2004_Sjj_ABS}. Note that the length $L_{1}$ and $L_{2}$ influences the $ABS$ energy via the two independent linear combination $\langle L \rangle$ and $\delta L$. Importantly, the term, $\mu \delta L/\hbar v_{F}$, is additive with  $\phi$ and  hence has exactly the same effect as $\phi$ - i.e.,  $\delta L \neq 0$ leads to  spontaneous Josephson effect, even when $\phi =0$. In the long junction limit, the ballistic region hosts multiple Andreev bound states ($\mathrm{ABS}$), of which only one pair is topological, crossing $E=0$ at $\theta = 2\mu \delta L/\hbar v_{F} - \phi = \pm \pi$. This can be confirmed by placing an impurity asymmetrically inside the junction (see Figure 1 in supplemental material). Unlike the short junction limit, where a single pair of topological $\mathrm{ABS}$ oscillates between the energy window $-\Delta_{0}$ to $\Delta_{0}$, in the long junction limit, the energy window of the oscillation of topological $\mathrm{ABS}$ is shortened by the factor $L_{SC}/\langle L \rangle$.

\vspace{0.2cm}

\noindent\emph{$Z_{2m}$ Parafermions:-}
\label{Parafermions}

\vspace{0.2cm}

Now we consider a set-up where the two quantum Hall liquids at filling fractions $\nu=1$ are replaced by $\nu=1/m$ and this  results in $4m\pi$ Josephson effect~\cite{cheng2015_FJJ,Cheng12,Jian_FJJ,CCkai_FJJ,frolov2020_FJJ,rokhinson2012_FJJ,badiane_FJJ}. As shown by Clarke \textit{et al.} \cite{Clarke2013_parafermions}, this is one of the simplest theoretical proposals for realizing parafermion zero modes.

At  the interface of the two quantum Hall liquids,
(shown in fig.~\ref{set-up}) 
the Hamiltonian for the gapless counter-propagating edge modes is given in bosonised form as 
\begin{equation}
    H_0 = \frac{mv_F}{4\pi} \int dx ~[(\partial_x \phi_{R})^2 + (\partial_x\phi_{L})^2]
\end{equation}
Here $v_F$ is the Fermi velocity and $m=1/\nu$ is the inverse of the filling fraction and  the  chiral fields $\phi_{R,L}$ satisfy
\begin{eqnarray}
\left[\phi_{R/L}(x), \phi_{R/L} (x')\right] &=& \pm
 i\frac{\pi}{m} \text{sgn}(x-x') \nonumber\\ 
 \left[ \phi_{R}(x), \phi_{L} (x')\right] &=&  i\frac{\pi}{m}
 \label{bosonic_comm}
\end{eqnarray}
These properties are sufficient to ensure  the proper anti-commutation relations for the fermion operators defined as
$\psi_{R/L} \sim e^{im\phi_{R/L}} $~\cite{haldane1981_bosonization,rao2002_bosonization,von1998_bosonization,giamarchi2003_bosonization,kane1992_bosonization}\footnote{While, for a single species of bosons,  our convention where the left and right-moving bosons have non-trivial commutation relations are sufficient to take care of Fermi statistics, the introduction of more species will require Klein factors  to ensure correct Fermi statistics between fermion operators}.

Next, we  briefly review the results of Lindner \emph{et al.} \cite{Lindner2012_parafermions} within our context. We imagine that the edge modes are fully gapped out by two alternating superconductors and ferromagnets (i.e., we imagine gapping out the free region in figure \ref{set-up}(a)  by  a ferromagnet $FM_1$.) The pairing due to the two superconductors  and the insulating gap induced by electron backscattering  are modelled by adding the appropriate cosine terms to the Hamiltonian, and the total Hamiltonian reads $H = H_{0} + H_{I}$, where
\begin{eqnarray}
H_I &=& \sum_{i=1,2}\left(\Delta_{i}\int_{SC_i}^{}dx~\text{cos}\left[m\left( \phi_R(x) + \phi_L(x)\right) \right] \right. \nonumber\\ 
&& \left. + \mathcal{M}_{i}\int_{FM_i}^{}dx~\text{cos}\left[m\left( \phi_R(x) - \phi_L(x)\right) \right]\right) 
\label{pair_hamiltonian}
\end{eqnarray}
The $SC/FM$ proximitized regions are characterized by  integer-valued charge/spin operators, called $\hat{Q}_j$ and $\hat{S}_j$ respectively.  More precisely, since the charge is defined modulo $2e$ in the $SC$ regions and the spin always changes in steps of $2$ (due to backscattering) in the $FM$ regions, the correct operators to describe the charge/spin in the $SC/FM$ regions are $e^{i\pi \hat{Q}_j}$ and $e^{i\pi\hat{S}_j}$. These operators are related to the bosonic fields as
\begin{eqnarray}
\hat{Q}_j &=& \int_{SC_j} dx~\frac{1}{2\pi}\partial_x(\phi_R - \phi_L)  \nonumber\\ 
\hat{S}_j &=& \int_{FM_j} dx~\frac{1}{2\pi} \partial_x(\phi_R + \phi_L)
\label{reln_charge_spin_boson}    
\end{eqnarray}
In the limit where $\Delta_{j}, \mathcal{M}_{j} \longrightarrow \infty$, the $\phi_{R}\pm \phi_{L}$ fields in equation \ref{pair_hamiltonian} are pinned to one of the $2m$ possible minima of the cosine, respectively. These minima are characterized by integer-valued operators $\hat{n}^{SC}_j$ in $SC_j$, and $\hat{n}^{FM}_j$ in $FM_j$. In the same limit, we can relate the operators $\hat{Q}_j, \hat{S}_j$ with $\hat{n}^{SC}_j, \hat{n}^{I}_j$ using Eq.(\ref{reln_charge_spin_boson}) giving us
\begin{equation}
    \hat{Q}_j/\hat{S}_j = \frac{1}{m}\left( \hat{n}^{FM/SC}_{j+1} - \hat{n}^{FM/SC}_{j} \right)
    \label{qs-n_reln}
\end{equation}
where the index $j$ is defined modulo 2.
Note that the SC/FM regions can exchange $1/m$ charges/spins with the bulk of the FQH systems. This means that the operators $e^{i\pi \hat{Q}_j}$ and $e^{i\pi\hat{S}_j}$ can have eigenvalues $e^{{i\pi q_j}/{m}}$ and, $e^{{i \pi s_j}/{m}}$ respectively, where $q_j, s_j \in \{0, 1, \dots  2m-1\} $. We now define the total charge and spin operators, $\hat{Q}_\text{tot}$, $\hat{S}_\text{tot}$, which satisfy the global constraint $e^{i\pi\hat{Q}_\text{tot}/\hat{S}_\text{tot}} = \prod_j e^{i\pi\hat{Q}_j/\hat{S}_j} = e^{i\pi(n_\uparrow \pm n_\downarrow)/m}$, where $n_{\uparrow/\downarrow}$ are the number of quasi particles in the spin up/down bulk FQH  regions. For a general $m$, the number of distinct values of $\lbrace n_{\uparrow} , n_{\downarrow}\rbrace$ consistent with the global constraints is $(2m)^{2}/2$~\cite{Lindner2012_parafermions}. Since, the two superconducting (ferromagnetic) segments are considered to be  parts of the same bulk superconductor (ferromagnet) (and the bulk SC is not assumed to be grounded), the total charge $q_{tot} = q_{1}+q_{2} $ and the total spin $s_{tot} = s_{1} + s_{2}$ of the system is conserved.

We hence label the ground state manifold by the eigenvalues of a \emph{complete set of mutually commuting operators.} The commutation relations detailed in the supplemental material show that our system hosts two such sets: $(e^{i\pi\hat{Q}_1}, e^{i\pi\hat{Q}_2}, \hat{S}_{tot}, H)$ and $(e^{i\pi\hat{S}_1}, e^{i\pi\hat{S}_2}, \hat{Q}_{tot}, H)$. The eigenvalues of both the sets of operators provide an  equivalent and a complete description of the ground state manifold of the system as long as the system is fully gapped by alternating superconductors and ferromagnets. The  degeneracy can then be counted by  the distinct set of eigenvalues of the operators in a particular basis subjected to global constraints. Note that for a fixed $\lbrace n_{\uparrow},n_{\downarrow}\rbrace$ sector, $s_{1}$ and $s_{2}$ are not independent. The commutation relations outlined in the supplemental material show that if $|s_{1},s_{2},q_{tot}\rangle$ is the eigenstate of the spin parity operator, $e^{i\pi \hat{S}_{i}}$, then so is $\left( e^{i\pi\hat{Q}_{1}}\right)^{k}|s_{1},s_{2},q_{tot}\rangle = | s_{1} + k,s_{2}-k,q_{tot} \rangle$, where $k \in \{0,\dots,2m-1 \}$. Hence, the ground state manifold is $2m$-fold degenerate for a fixed $\lbrace n_{\uparrow},n_{\downarrow}\rbrace$. {Counting all} possible values of $\lbrace n_{\uparrow},n_{\downarrow}\rbrace$ gives the dimension of the ground state Hilbert space to be $(2m)^{3}/2$. The same set of arguments above can be repeated for the states labelled by $\ket{q_1,q_2,s_\text{tot}}$ {to obtain the same results}.

Now, let us  remove one of the insulating gaps, by taking $\mathcal{M}_{1} \rightarrow 0$. This 
leads to the realization of the ballistic Josephson junction setup as given in Fig.\ref{set-up}(a).  For fixed $\lbrace n_{\uparrow},n_{\downarrow} \rbrace$, the $2m$ states, which were degenerate ground states in the large $\mathcal{M}_{1}$ limit, now move away from zero energy and are no longer
degenerate. The actual splitting of the energy depends on the various parameters - $\phi$, $\delta L$ and
$\langle L \rangle$.   Furthermore, as  $\mathcal{M}_{1}\rightarrow 0$, {the two superconductors are connected by the junction and }the charge parity operators, $e^{i\pi \hat{Q}_{i}}$, no longer commute with the Hamiltonian, that is $\left[ e^{i\pi\hat{Q}_{i}},H \right] \neq 0$.  However, the other set of operators, $\hat{S}_{1}$, $\hat{S}_{2}$ and $\hat{Q}_{tot}$, still commute with the Hamiltonian. This means that
rather than the basis, $| q_{1},q_{2},s_{tot} \rangle$, we should use the eigenvalues of the set of mutually commuting operators, $\hat{S}_{1}$, $\hat{S}_{2}$, $\hat{Q}_{tot}$ to label the {states} as $|\bar{s}_{1},\bar{s}_{2},q_{tot} \rangle$. Note that we now label the eigenstates with the eigenvalues $\bar{s}_{j}$ of the operator $\hat{S}_{j}$ rather than those of {the spin parity} $e^{i\pi\hat{S}_{j}}$ since removing $FM_1$ precludes backscattering between the edges.
We will show later that the energy eigenvalue depends only  on the spin in  the ballistic JJ region and is given by $H |\bar{s}_{1},\bar{s}_{2},q_{tot}\rangle = E(\bar{s}_{1})|\bar{s}_{1},\bar{s}_{2},q_{tot}\rangle$ (see Eq.~\ref{Eq:H_eff}). Thus, the $2m$ ground states, which were degenerate at $E=0$ in the $\mathcal{M}_{1} \rightarrow \infty$ limit, are now at different energies $E(\bar{s}_{1})$ for the $2m$ possible values of $\bar{s}_{1}$. As we change the phase factor $\theta = 2\mu \delta L/\hbar v_{F} - \phi$, the eigenvalues oscillate and cross each other.  

{As was shown earlier,} the  effective theory of the Josephson junction between $SC_1$ and $SC_2$, when $L_1=L_2$, exhibits the  Josephson effect with a  periodicity  $4\pi m$\cite{Clarke2013_parafermions}.
For different lengths, we first note that the ABS spectrum derived in the supplemental essentially used the fact that particles and holes transform back into themselves after two consecutive Andreev reflections, having traversed a path of length $L_1+L_2$. Thus, the spectrum includes the effect of the Andreev reflections as well as the dynamical phases. In terms of twisted boundary conditions, this translates to 
\begin{eqnarray}
\psi_{R}(x+L_1+L_2) &=& e^{-2i\theta}e^{i\left(k_{e}L_{1}-k_{h}L_{2} + \phi_{2}-\phi_{1}\right)} \psi_{R}(x) \nonumber\\
&\equiv& e^{i\sigma} \psi_R (x)
\label{Eq:BC}
\end{eqnarray}
where $\sigma/2 = -\cos^{-1}\left(\frac{E}{\Delta_{0} } \right) +  \frac{E\langle L \rangle}{\hbar v_F} \pm \left( \frac{\mu \delta L}{\hbar v_F} - \frac{\phi}{2} \right)$ represents all the phases accumulated by an electron when it traverses the loop defined by Andreev reflections between the two ends of the junction, and $\phi \equiv \phi_2 - \phi_1 $.
{We then note that in terms of the bosonised Hamiltonian, this translates into the superconducting coupling between the two counter-propagating edge states of the following form}:
\begin{eqnarray}
{H_{SC}} &=& -\Delta_{0}\left(\int_{-l_{SC}}^0 dx ~\text{cos}\Big[m\big(\phi_R(x) + \phi_L(x) \big)\Big]\right.\nonumber\\ 
&& \left. +\int_{L_1}^{L_{1}+l_{SC}} dx ~\text{cos}\Big[m\big(\phi_R(x) + \phi_{L}(x-2\delta L)\big) + \sigma\Big]\right) \nonumber\\ 
\end{eqnarray}
where $l_{SC}$ is the length of the superconducting regions. Note also that all the phases {($\sigma$)} accumulated in traversing the loop between the two superconductors have been plugged into the second superconductor using gauge freedom. $\Delta_0$ is the magnitude of the superconducting pairing. 

Thus, the total Hamiltonian is given by ${H = H_{0} + H_{SC}}$. 
In the $\Delta_0 \rightarrow \infty$ limit, as remarked earlier, the field $\phi_R + \phi_L$ is confined to the minima of the cosine potential and $E\ll\Delta_0$, giving us
$\sigma = 2\pi \pm (\frac{2\mu \delta L}{\hbar v_F} - \phi)$, resulting in the following boundary conditions for the finite-length chiral Luttinger liquids in the junction between the two superconductors: 
\begin{equation}
\begin{split}
    \phi_R(0) + \phi_L(0) &= 0 \\
    \phi_R(L_1) + \phi_L(L_2) &= 2~\left(\text{mod}\Big[ \frac{\pi}{m}\Big( \hat{n}^{{SC}}_2 - \frac{\sigma}{2\pi} \Big)  , 2\pi \Big] - \pi\right)\\
    &\equiv 2\hat{\eta}
\end{split} \label{bosonicbc1}
\end{equation}
where $\hat{n}^{SC}_2$ is an integer-valued operator corresponding to the pinned minimum of the fields at the right superconductor such that it can assume $2m$ values, $n^{SC}_{2} \in \lbrace 0, 2m-1\rbrace$. $\hat{n}^{SC}_{1}$ can be taken as zero without loss of generality. The modulus  is necessary to ensure the compactness of the finite-length bosonic fields. It is interesting to note from equation \ref{reln_charge_spin_boson} that {$\hat{\eta}/\pi$} is nothing but the spin $\hat{S}_1$ of the junction. The effective Hamiltonian for the ballistic junction between the two superconductors is given by:
\begin{equation}
   H_\text{eff} =  \frac{mv_F}{4\pi} \int_{-L_2}^{L_1}dx\left(\partial_x\phi_R(x)\right)^2
\end{equation}
where,  $ \phi_{R}(x,t)$ is given by \cite{udit&yuval_parafermion} {(see also supplemental material for more details)}
\begin{align}
    &\phi_{R}(x) = \frac{2\hat{\eta} }{L_{1} + L_{2}}(x-L_{1}) +\hat{\chi}
\nonumber \\
&+ \frac{1}{\sqrt{m}}\sum_{k>0}\left( \hat{a}_{k}e^{\frac{2\pi i k}{L_{1} + L_{2}}(x-L_{1})} + \hat{a}_{k}^{\dagger}e^{-\frac{2\pi i k}{L_{1} + L_{2}}(x-L_{1})}\right)
\end{align}
with $\phi_L(x) = -\phi_R(-x)$ and $\left[ n_2^{SC},\hat{\chi}\right] = i$, such that equations \ref{bosonic_comm} and \ref{bosonicbc1} are satisfied. This diagonalizes the effective Hamiltonian, giving us
\begin{equation}
\begin{split}
    H_{\text{eff}} = \frac{mv_F}{\pi(L_1 + L_2)}\hat{\eta}^2 + \sum_{k>0}\frac{2\pi k v_F}{L_1+L_2}\Big(a^{\dagger}_{k}a_{k} + \frac{1}{2}  \Big).
\end{split}
\label{Eq:H_eff}
\end{equation}

In Eq.~\ref{Eq:H_eff}, the first term carries the dependence of energy on the SC phase difference $\phi$ and on the additional phase arising due to length difference in the two chiral modes. Importantly, we note that the $\left[{H}_{\text{eff}}, \hat{n}^{SC}_{2}\right] = 0$ and as result { $\hat{n}^{SC}_{2}$}, is a conserved quantity. For a fixed eigenvalue of the $\hat{n}^{SC}_{2}$ operator, the energy is $4m\pi$ periodic in $\theta = 2\mu\delta L/\hbar v_{F} -\phi$. The Josephson current across the ballistic region, $I_{\theta} \propto d\langle H \rangle/d \theta$, also shows the $4m\pi$ periodicity in $\theta$.

\noindent\emph{ Discussion and Conclusion:} \label{discussion}

\begin{figure}
\centering
\includegraphics[scale=0.10]{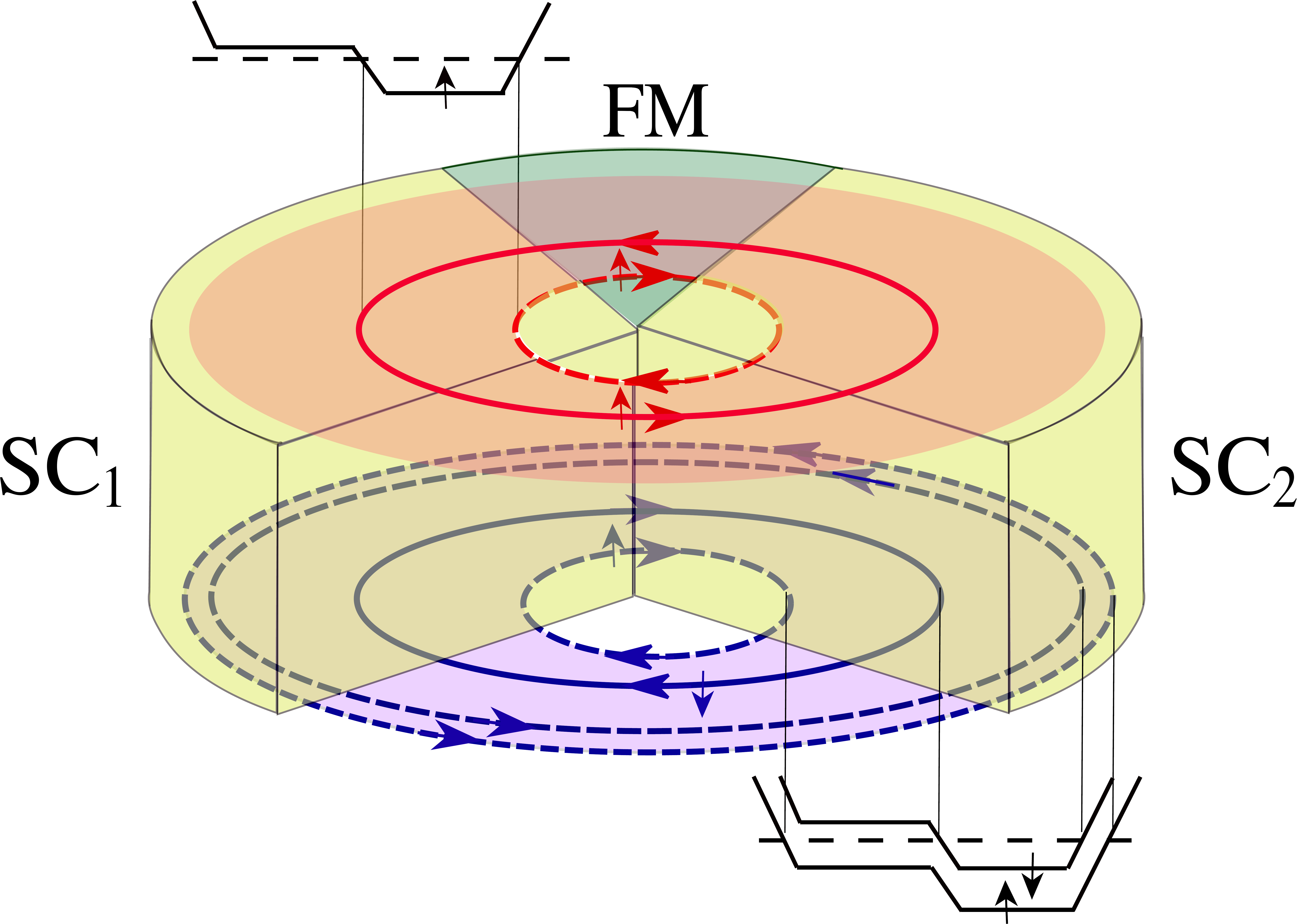}
\caption{A proposed set-up to realize the fractional Josephson effect in a bilayer FQH system, {with the top layer at $\nu = 1/m$ and the bottom layer at $\nu = 1+1/m$.} The Landau levels are manipulated using appropriate gating such that two counter-propagating chiral states with opposite spins are brought together. The chiral states at the middle of the sample (shown in red and blue solid lines) are of importance to realize Josephson junction geometry. These chiral states are proximitized by two superconductors, $SC_{1}$ and $SC_{2}$, and a ferromagnet ($FM$) at the back. The length of  the individual counter-propagating chiral states, in the ballistic region, can be altered using the external gates, which can drive the fractional Josephson current and show $4\pi m$ periodicity. Inconsequential chiral edge states are shown with broken lines (red and blue) in the two layers.}
\label{set-up_moty}
\end{figure}

The main focus of this paper has been to show that allowing the length  of the counter-propagating chiral edge states, belonging to two FQH systems, to be different, introduces a new experimental knob on equal footing with SC phase bias, hence leading to spontaneous fractional Josephson effect. We have first demonstrated the feasibility in a $\nu=1$ quantum Hall set-up where the Andreev modes can be computed exactly and shown how the length difference can lead to a spontaneous Josephson current. We have then extended our study to a $\nu=1/m$ set-up with $Z_{2m}$ parafermion modes between the superconductors leading to
a spontaneous $4\pi m$ Josephson effect tunable by the difference in the lengths of the two edges. Such a finding may be of importance because it provides an extra handle on the Josephson {current}, controllable by electrical means, to probe parafermions. For  $v_F \sim 10^4~m/s$ and $\mu \sim 10~ meV$ \cite{exp_parameters1, exp_parameters2}, change in $\delta L$ required to access the $4\pi m$ Josephson effect turns out to be a few $\mu m$ in conventional 2DEG systems, making it experimentally accessible by current standards.

To this end, we propose a setup to realize the spontaneous fractional Josephson current in a 2DEG embedded in a double quantum well tuned to {two different} FQH states (see fig.~\ref{set-up_moty}). This setup is inspired by the experiment in \onlinecite{ronen2018_HelEdge}. We get two counter-propagating chiral edge states at the center of two FQH system with opposite spins, which can be proximitized by the SC and FM as shown in fig.~\ref{set-up_moty}. The external gates used to manipulate the Landau levels can also be used to displace the chiral edge state at the center of the sample by changing the gate strength (voltage) and hence in principle, changing the length of the chiral edge state in the ballistic region. This external control on the length of the chiral edge state gives an experimental handle to realize the spontaneous fractional JJ effect.

\begin{acknowledgments}
We acknowledge early collaboration with Krashna Mohan Tripathi and
wish to thank him for many useful discussions. 
A.R. acknowledges University Grants Commission, India, for support in the form of a fellowship. S.D. would like to acknowledge the MATRICS grant (Grant No. MTR/ 2019/001 043) from the Science and Engineering Research Board (SERB) for funding. S.D. also acknowledges warm hospitality from ICTS during the final stages of writing the draft. 
K.I. thanks the ICTS - Long Term Visiting Students Program 2021.
\end{acknowledgments}

\vspace{+0.2cm}
\emph{Author contribution:-}
The first two authors, K.I.  and A.R. have contributed equally to this work. \\ 

\bibliography{citations.bib}
\end{document}


\title{Supplemental material for ``Spontaneous fractional Josephson current  from parafermions"}
\author{Kishore Iyer}
\affiliation{Aix Marseille Univ, Université de Toulon, CNRS, CPT, Marseille, France}
\affiliation{International Centre for Theoretical Sciences, Tata Institute
of Fundamental Research, Bengaluru 560089, India}
\author{Amulya Ratnakar}
\affiliation{Department of Physics, Indian Institute of Science Education and Research (IISER) Kolkata, Mohanpur - 741246, West Bengal, India}
\email{ORCID ID: 0000-0002-0025-9552}
\author {Aabir Mukhopadyay}
\affiliation{Department of Physics, Indian Institute of Science Education and Research (IISER) Kolkata, Mohanpur - 741246, West Bengal, India}
\email{ORCID ID: 0000-0001-6465-2727}
\author{Sumathi Rao}
\affiliation{International Centre for Theoretical Sciences, Tata Institute
of Fundamental Research, Bengaluru 560089, India}
\author{Sourin Das}
\affiliation{Department of Physics, Indian Institute of Science Education and Research (IISER) Kolkata, Mohanpur - 741246, West Bengal, India}

\maketitle

\numberwithin{equation}{section}

\begin{widetext}

\section{Andreev bound states}\label{ABS_spectrum_app}

We start with the Hamiltonian of two counter-propagating fermionic edges

\begin{equation}
    H = (-i\hbar v_F \partial_x\sigma_z -\mu )\tau_z + \Delta (x)(\text{cos}\phi_r \tau_x - \text{sin}\phi_r \tau_y) \label{eqone}
\end{equation}
which has been written in the Nambu basis $ \big( \psi_R, \psi_L, \psi_L^\dagger, -\psi_R^\dagger \big) $. We will define the reflection and transmission matrices for this set-up following the method in \cite{Kundu2010}. At the two NS junctions, the particles undergo reflections given by:
\begin{align}
\begin{split}
 \psi_R(x = L_1) &\longrightarrow r^1_{Ahe}\psi_L^\dagger(x=L_2) + r^1_{Nee}\psi_L(x=L_2) \\
 \psi_L^\dagger(x=0) &\longrightarrow r^1_{Aeh}\psi_R(x=0) + r^1_{Nhh}\psi_R^\dagger(x=0) \\
 \psi_R^\dagger(x=L_1) &\longrightarrow r^2_{Ahe}\psi_L(x=L_2) + r^2_{Nhh}\psi_L^\dagger(x=L_2) \\
 \psi_L(x=0) &\longrightarrow r^2_{Ahe}\psi_R^\dagger(x=0) + r^2_{Nee}\psi_R(x=0)
\end{split}
\end{align}
Assuming perfect Andreev reflection, we have $r^1_{Nee} = r^1_{Nhh} = r^2_{Nee} = r^2_{Nhh} = 0$, and $r^1_{Ahe} = e^{-i\theta}e^{i\phi_1}$, $r^1_{Aeh} = e^{-i\theta}e^{-i\phi_2},
r^2_{Ahe} = e^{-i\theta}e^{-i\phi_1}$,
$r^2_{Ahe} = e^{-i\theta}e^{i\phi_2}$. 
With these reflection elements, we define the reflection matrix which acts at both the NS junctions as:
\begin{align}
    \mathcal{R} = e^{-i\theta}\begin{bmatrix}
    0 & 0 & e^{i\phi_1} & 0\\
    0 & 0 & 0 & e^{i\phi_2} \\
    e^{-i\phi_2} & 0 & 0 & 0 \\
    0 & e^{-i\phi_1} & 0 & 0 \\
    \end{bmatrix}
\end{align}
where $\theta = \cos^{-1}\left(\frac{E}{\Delta}\right)$ and $E$ is the energy of the particles. The translation matrix which translates a right moving particle/hole over a length $L_1$ and a left moving particle/hole over a length $L_2$ is given by

\begin{equation}
    \mathcal{T} = \text{diag}\begin{bmatrix} e^{ik_eL_1} & e^{ik_eL_2} & e^{-ik_h L_2} & e^{-ik_h L_1},
\end{bmatrix}
\end{equation}
where, $k_{e/h} = \frac{\mu \pm E}{\hbar v_F}$. Two consecutive Andreev reflections occur over the length of $L_1+L_2$, such that,
\begin{eqnarray}
\mathcal{R}\mathcal{T}\mathcal{R}\mathcal{T} \psi(x) &=& \psi(x+L_{1}+L_{2})
\label{Eq:Mat_form_AR_loop}
\end{eqnarray}

where,

\begin{eqnarray}
\mathcal{R}\mathcal{T}\mathcal{R}\mathcal{T} &=& e^{-i2\theta}\begin{pmatrix}
             e^{i(k_{e}L_{1} -  k_{h}L_{2} + \phi_{1}-\phi_{2})} & 0 & 0 & 0 \\
             0 & e^{i(k_{e}L_{2} -  k_{h}L_{1} + \phi_{2}-\phi_{1})} & 0 & 0 \\
             0 & 0 & e^{i(k_{e}L_{1} -  k_{h}L_{2} + \phi_{1}-\phi_{2})} & 0 \\
             0 & 0 & 0 & e^{i(k_{e}L_{2} -  k_{h}L_{1} + \phi_{2}-\phi_{1})}
             \end{pmatrix}.
\end{eqnarray}
To obtain the ABS spectrum we recognize the fact that particles must come back to themselves after two consecutive Andreev reflections. This gives us the following determinant condition:

\begin{equation}
    \text{Det} \big[ \mathcal{I} - \mathcal{R}\mathcal{T}\mathcal{R}\mathcal{T}  \big] = 0
\end{equation}
where $\mathcal{I}$ is the identity matrix. Solving this determinant equation for the energy $E$, one obtains the following transcendental equation.
\begin{equation}
    E = \pm\Delta_{0}~\text{cos}\Bigg[ \frac{E\langle L\rangle}{\Delta_{0} L_{SC}} \pm \left(\frac{\mu\delta L}{\hbar v_F} -\frac{\phi}{2}  \right) \Bigg]
    \label{Eq:ABS_Clean}
\end{equation}

where $\langle L\rangle = \frac{L_1 + L_2}{2}$, $\delta L = \frac{L_1 - L_2}{2}$, $\phi = \phi_1- \phi_2$ and $L_{SC} = \hbar v_{F}/\Delta_{0}$.

\section{The false Majorana states}
\label{Appendix:B}

We now introduce a scatterer in the ballistic region as shown in fig.~\ref{fig:ABS_scatterer}. 
The effect of this scatterer on electrons and holes impinging on it is given by the matrix

\begin{equation}
\mathcal{S} = \begin{pmatrix}
			 t & i\sqrt{1-t^{2}} & 0 & 0 \\
			 i\sqrt{1-t^{2}} & t & 0 & 0 \\
			 0 & 0 & t & i\sqrt{1-t^{2}} \\
			 0 & 0 & i\sqrt{1-t^{2}} & t	
			 \end{pmatrix}
\end{equation}
Let the right moving edge be defined over the $1D$ space $x \in [-L_1/2, L_1/2]$ and the left moving edge be defined over $x \in [-L_2/2, L_2/2].$ Then placing the scatterer symmetrically in the ballistic junction amounts to placing the scatterer at $x=0$. To break the parity symmetry, we place the scatterer asymmetrically in the junction, at $x=-\alpha$. We now define two translation matrices, namely $\mathcal{T}_1$ and $\mathcal{T}_{2}$, which takes the particle/hole towards and away from the scatterer.
\begin{eqnarray}
\mathcal{T}_1 &=& \mathrm{diag}\left[ e^{ ik_{e}\left(\frac{L_{1}}{2} -\alpha\right)}, e^{ ik_{e}\left(\frac{L_{2}}{2} +\alpha\right)},  e^{ -ik_{h}\left(\frac{L_{2}}{2} + \alpha\right)}, e^{ -ik_{h}\left(\frac{L_{1}}{2} -\alpha\right)}\right] \nonumber\\
\mathcal{T}_2 &=& \mathrm{diag}\left[ e^{ ik_{e}\left(\frac{L_{1}}{2} +\alpha\right)}, e^{ ik_{e}\left(\frac{L_{2}}{2} -\alpha\right)},  e^{ -ik_{h}\left(\frac{L_{2}}{2} - \alpha\right)}, e^{ -ik_{h}\left(\frac{L_{1}}{2} +\alpha\right)}\right] \nonumber\\
\end{eqnarray}

\begin{figure}
    \centering
    \includegraphics[scale=0.2]{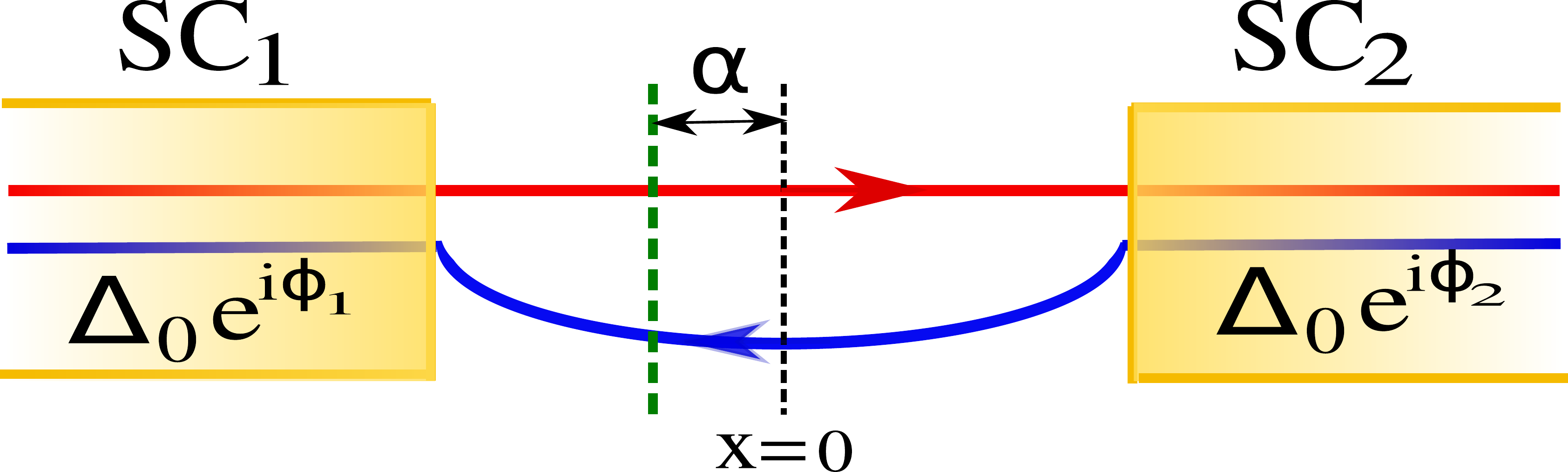}
    \caption{The figure shows a Josephson junction setup consisting of two counter-propagating edge states corresponding to $\nu={1}$ proximitised by two superconductors $SC_{1}$ and $SC_{2}$. $\Delta_{0}$ is the superconducting gap and $\phi_{i}$ is the superconducting phase corresponding to $SC_{i}$. The right and left moving edge (in red and blue) is taken to be of length $L_{1}$ and $L_{2}$, respectively. The right(left) moving edges are defined over $x \in [-L_1/2, L_1/2]$ $([-L_2/2, L_2/2])$ about the origin $x=0$. The scatterer is placed at a distance $\alpha$ away from the origin. }
    \label{fig:ABS_scatterer}
\end{figure}
The ABS spectrum can be found as earlier, using the argument that the fermionic field returns to itself after a cycle of length $L_{1}+L_{2}$ after consecutive Andreev reflections from both boundaries. This, in the matrix formulation reduces to 
\begin{eqnarray}
\mathcal{R}\mathcal{T}_2\mathcal{S}\mathcal{T}_1\mathcal{R}\mathcal{T}_2\mathcal{S}\mathcal{T}_1\psi(x) &=& \mathcal{I}~\psi(x+L_{1}+L_{2}) 
\label{ABS_defn_asymm_scat}
\end{eqnarray}
\begin{eqnarray}
\mathrm{Det}\left[\mathcal{I}- \mathcal{R}\mathcal{T}_2\mathcal{S}\mathcal{T}_1\mathcal{R}\mathcal{T}_2\mathcal{S}\mathcal{T}_1\right] &=& 0
\label{Eq: Quantization_Condition}
\end{eqnarray}
and from Eq.~\ref{Eq: Quantization_Condition}, we get the quantization condition as

\begin{equation}
(1-t^{2})\sin^{2}(\alpha (k_e-k_h)) +  t^{2} \cos^{2}\left(\frac{\mu\delta L}{\hbar v_{F}} - \frac{\phi}{2}\right)
= \cos^{2}\left(\frac{E \langle L \rangle}{\hbar v_{F}} - \theta \right)
\end{equation}
ABS spectrum is given by the self-consistent equation
\begin{equation}
E = \pm \Delta_{0}\cos \left[ \frac{E \langle L \rangle}{\hbar v_{F}}  \pm \right. 
\left. \cos^{-1}\left(\sqrt{(1-t^{2})\sin^{2}\left(\frac{2 \alpha E}{\hbar v_{F}}\right) + t^{2}\cos^{2}\left(\frac{\mu \delta L}{\hbar v_{F}} - \frac{\phi}{2}\right)}\right)\right]
\end{equation}

\begin{figure}
\includegraphics[scale=0.27]{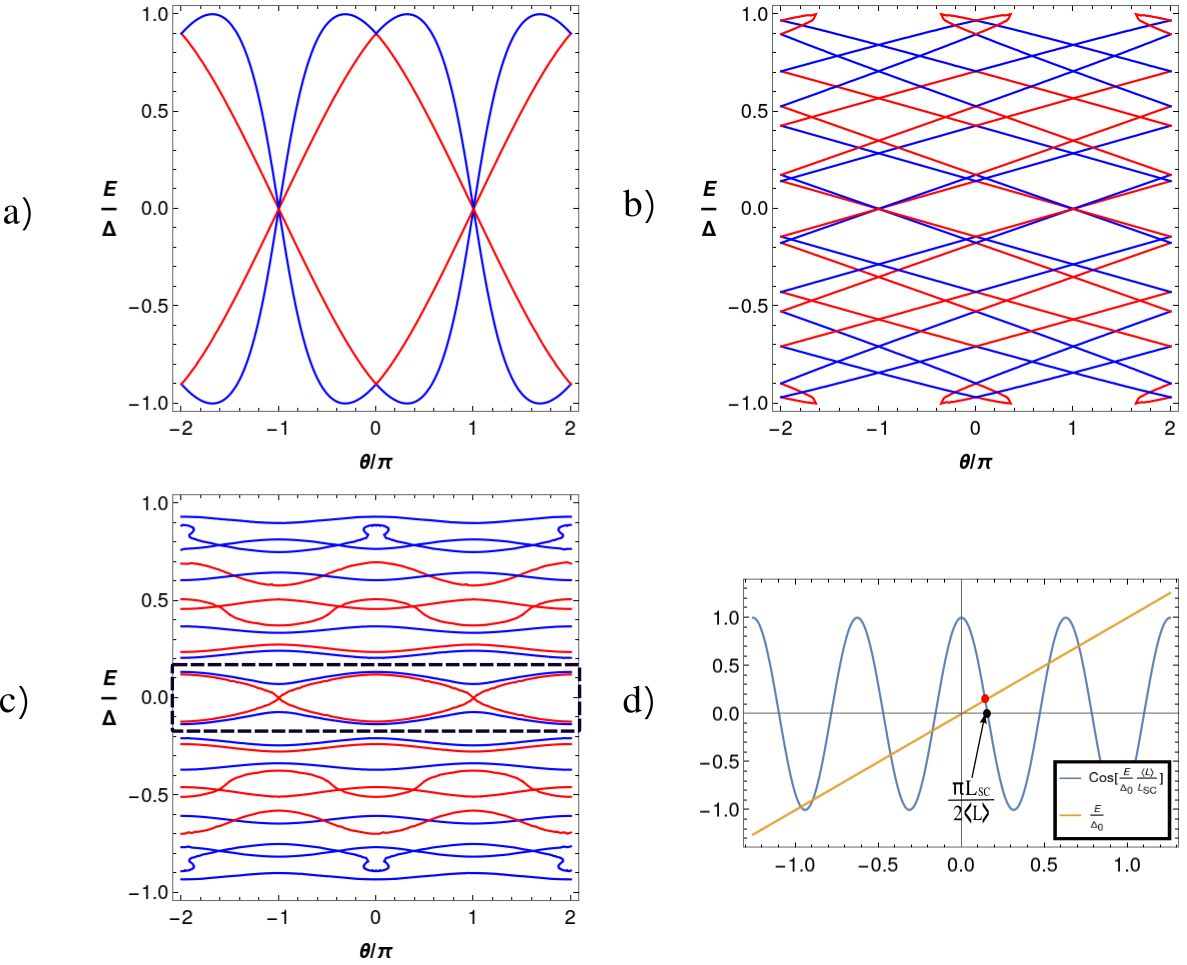}
\caption{Andreev bound states (ABS) are plotted as a function of {$\theta \equiv -\phi + 2\mu \delta L/\hbar v_{F}$}. ABS is plotted for a) small junction limit with $\langle L\rangle = 0.5 L_{SC}$, $\alpha = 0$ and $t =1$, b) long junction limit with $\langle L\rangle = 10 L_{SC}$,  $\alpha = 0$ and $t = 1$, c) long junction limit with $\langle L\rangle = 10 L_{SC}$, $\alpha = 8 L_{SC}$ and $t = 0.75$. Topological ABS (red) is shown in the dashed box. Fig. d) shows the self-consistent solution of Eq.~\ref{Eq:ABS_Clean} for $\langle L\rangle = 10 L_{SC}$ at $\theta = 0$. Red dot denotes the doubly degenerate smallest solution of Eq.~\ref{Eq:ABS_Clean}. }
\label{Fig_ABS}
\end{figure}

For $t=1$, the scattering matrix is fully transmitting with $\mathbb{S} = \mathbb{I}_{4\times 4}$. At  $0\leq t<1$ and non-zero $\alpha$, we see that two of the four states which were at zero energy at  $\frac{2\mu \delta L}{\hbar v_{F}} - \phi = \pm\pi$ gap out. For convenience in depicting the plots, we define
$\frac{2\mu \delta L}{\hbar v_{F}} - \phi \equiv \theta$. The length scale can be normalized with respect to superconducting phase coherence length $L_{S} = \hbar v_{F}/\Delta_{0}$. In Fig.~\ref{Fig_ABS}b, we note that, for $t=1$, there are multiple ABS, but there are only two pairs of ABS that crosses $E=0$ at the Dirac points, $\theta = \pm \pi$. In the presence of a scatterer placed asymmetrically (Fig.~\ref{Fig_ABS}c), two of the four ABS no longer cross the Dirac points while the other two ABS are topologically protected. These topologically protected ABS are the Majorana zero modes of the system. For the fully transmitting case ($t=1$) in the short junction limit ($L_1, L_2 << L_{S}$), we have $ E = \pm \Delta_{0} \cos\left(\frac{\phi}{2}\right)$ and we get two sets of doubly degenerate ABS as in \cite{fu2009_majorana}.

\section{Commutation relations}\label{Commutation_relation}

Here we briefly outline commutation relations following \cite{Lindner12_parafermion}. Let $n_{\uparrow/ \downarrow}$ be the total number of quasi particles in the bulk of  the spin up/down FQH liquids. We define the total charge and total spin operators which satisfy a constraint imposed by the bulk 
\begin{equation}
\begin{split}
    e^{i\pi\hat{Q}_\text{tot}} = \prod_j e^{i\pi\hat{Q}_j} = e^{i\pi(n_\uparrow + n_\downarrow)/m} \\
    e^{i\pi\hat{S}_\text{tot}} = \prod_j e^{i\pi\hat{S}_j} = e^{i\pi(n_\uparrow - n_\downarrow)/m}.
    \label{qpconstraint}
\end{split}
\end{equation}
Here, $q_{tot}$ and $s_{tot}$, the eigenvalues of $\hat{Q}_{tot}$ and $\hat{S}_{tot}$ respectively, are constrained to be even or odd simultaneously, giving us $2m^{2}$ distinct  $\lbrace n_{\uparrow},n_{\downarrow} \rbrace$ pairs corresponding to different bulk constraints.

The charge and the spin operators satisfy the  commutation relations $\left[\hat{Q}_{i},\hat{Q}_{j}\right] = \left[\hat{S}_{i},\hat{S}_{j}\right] = 0$. Then the appropriate parity operators  $e^{i\pi\hat{Q}_{i}}$ and $e^{i\pi \hat{S}_{i}}$  satisfy

\begin{eqnarray}
\left[ e^{i\pi \hat{Q}_j}, H \right] &=& 0 = \left[e^{i\pi \hat{S}_j}, H \right] \nonumber\\
\left[e^{i\pi \hat{Q}_j}, e^{i\pi \hat{S}_\text{tot}} \right] &=& 0 = \left[e^{i\pi \hat{S}_j}, e^{i\pi \hat{Q}_\text{tot}} \right] \nonumber\\
e^{i\pi \hat{Q}_i} e^{i\pi \hat{S}_j} &=&  e^{-\frac{i \pi}{m}\left(\delta_{i,j+1}-\delta_{i,j}\right)}e^{i\pi \hat{S}_j} e^{i\pi \hat{Q}_i}.
\label{Eq:Ady_stern_commutation}
\end{eqnarray}

\section{Bosonization details}

Fractional quantum Hall edge states are modelled by chiral bosonic fields $\phi_{R/L}.$ The right/left-moving fermions on the edges are then given by
\begin{equation}
    \psi_{R/L}(x) = \frac{1}{\sqrt{2\pi a}}e^{im\phi_{R/L}}
\end{equation}
where $m$ is the inverse filling fraction and $a$ is a cutoff parameter. To ensure the correct fermionic anticommutators, the chiral bosonic fields must obey the following commutation relations
\begin{subequations}
\begin{align}
[\phi_{R/L}(x), \phi_{R/L} (x')] = \pm i\frac{\pi}{m} sgn(x-x')  \label{comm1} \\
[\phi_{R}(x), \phi_{L} (x')] =  i\frac{\pi}{m} \label{comm2}
\end{align} 
\end{subequations}
Here, Eq.\ref{comm1} ensures the correct anticommutator for fermions on the same chiral edge while Eq.\ref{comm2} ensures the same for fermions on different edges. These commutators are sufficient to ensure correct fermionic behavior as long as we consider only two chiral edges. For a geometry involving three or more edges one would need to introduce Klein factors \cite{Guyon02}.

The Hamiltonian for the system we are looking at is given by $H = H_0 + H_I$ where $H_0$ is the bosonized Hamiltonian of the counter-propagating FQH edges modelled by chiral Luttinger liquids
\begin{equation}
    H_0 = \frac{mv_F}{4\pi} \int dx ~[(\partial_x \phi_{R})^2 + (\partial_x\phi_{L})^2].
\end{equation}
and $H_I$ models the superconducting pairing between the two FQH edges.
\begin{equation}
\begin{split}
    H_I = -\int_{-\infty}^0 dx ~\Delta_0~\text{cos}\Big[m\big(\phi_R(x) + \phi_L(x) \big)\Big] \\ 
    - \int_{L_1}^\infty dx~\Delta_0~ \text{cos}\Big[m\big(\phi_R(x) + \phi_L(x+L_2-L_1)\big) + \sigma\Big]
\end{split} \label{intham}
\end{equation}
Being interested in the Josephson periodicity of this set-up, we work in the low-energy/strong-coupling limit, considering $E \ll \Delta_0.$ In this limit, we only look at the island between the two superconductors. One expects this to be modelled by the following effective Hamiltonian
\begin{equation}
    H_{\text{eff}} = \frac{mv_F}{4\pi} \bigg[ \int_0^{L_1}dx~(\partial_x \phi_R)^2 + \int_0^{L_2}dx~(\partial_x \phi_L)^2 \bigg] 
    \label{effectiveHamiltonian1}
\end{equation}
which along with the boundary conditions encapsulating the effect of the superconducting pairing on the chiral bosonic fields
\begin{equation}
\begin{split}
    \phi_R(0) + \phi_L(0) &= 0 \\
    \phi_R(L_1) + \phi_L(L_2) &= 2~\text{mod}\Big[ \frac{\pi}{m}\Big( \hat{n}^{{SC}}_2 - \frac{\sigma}{2\pi} \Big)  , 2\pi \Big] \\
    &\equiv 2\hat{\eta}
\end{split} \label{bosonicbc}
\end{equation}
suggests the following mode expansion to diagonalize the Hamiltonian
\begin{equation}
\begin{split}
    \phi_R(x) = \hat{\eta} \frac{x}{L_1} + \chi + \sum_{k>0}\frac{1}{\sqrt{mk}}\left[ e^{\frac{ikx}{L_1}}a_k + e^{-\frac{ikx}{L_1}}a_k^\dagger \right] \\
        \phi_L(x) = \hat{\eta} \frac{x}{L_2} - \chi + \sum_{k>0}\frac{1}{\sqrt{mk}}\left[ e^{\frac{ikx}{L_2}}a_k + e^{-\frac{ikx}{L_2}}a_k^\dagger \right].
\end{split}
\end{equation}
However, {an explicit computation of  the} effective Hamiltonian  reveals two things: (i) The modes appear to be quantized over over two different lengths $L_1$ and $L_2$ (ii) There are leftover linear $a_k, a_k^\dagger $ terms, rendering the model unphysical. What went wrong here is precisely that we failed to account for the correct quantization of the bosonic modes in the junction region. The bosonic fields are spread over the entire junction (of length $L_1 + L_2$), and not over the individual chiral Luttinger liquids (of lengths $L_1$ and $L_2$ respectively) suggesting that they should now be quantized over a length $L_1+L_2$. This can be corrected by suitably modifying the effective Hamiltonian, which turns out to be
\begin{equation}
\begin{split}
        H_{\text{eff}} = \frac{mv_F}{4\pi} \bigg[ \frac{2L_1}{L_1+L_2}\int_0^{L_1}dx~(\partial_x \phi_R)^2 \\ + \frac{2L_2}{L_1+L_2}\int_0^{L_2}dx~(\partial_x \phi_L)^2 \bigg]   
    \label{effectiveHamiltonian2}
\end{split}
\end{equation}
Now we see that the above mode expansion diagonalizes the Hamiltonian, giving us the energy spectrum
\begin{equation}
    H = \frac{mv_F}{\pi(L_1+L_2)}\hat{\eta}^2 + \sum_{k>0} \frac{2\pi v_F k}{L_1 + L_2}\left(a_k^\dagger a_k + \frac{1}{2}\right)
    \label{diag_eff_H}    
\end{equation}
from which it is clear that the quantization is over the length $L_1 + L_2$. The fact that bosonic fields are quantized over the length $L_1+L_2$ instead of the individual lengths motivates us to think of this system as a ring of length $L_1 + L_2$ described by a single global chiral bosonic field $\tilde{\phi}$ related to the original $\phi_R, \phi_L$ fields as
\begin{equation}
    \begin{split}
        \phi_R(x) &= \tilde{\phi}\left(\frac{2L_1}{L_1+L_2}x\right) \\
        \phi_L(x) &= \tilde{\phi}\left(-\frac{2L_2}{L_1+L_2}x\right)
    \end{split}
\end{equation}
Given that the original fields $\phi_R$ and $\phi_L$ are defined over the $1D$ spaces $[0,L_1]$ and $[0,L_2]$, respectively, $\tilde{\phi}$ is defined over the $1D$ space $[-\frac{L_1+L_2}{2},\frac{L_1+L_2}{2}]$. Starting from equation \ref{effectiveHamiltonian1} and correctly applying the transformations, we get the effective Hamiltonian in terms of the global field $\tilde{\phi}$
\begin{equation}
    H_\text{eff} = \frac{mv_F}{4\pi}\int_{-\frac{L_1+L_2}{2}}^\frac{L_1+L_2}{2}dx \left(\partial_x \tilde{\phi}(x) \right)^2
\end{equation}
 which is diagonalized by the mode expansion
 \begin{equation}
    \tilde{\phi}(x) = \frac{2\hat{\eta}}{L_1+L_2}x + \hat{\chi} + \sum_{k>0} \frac{1}{\sqrt{mk}}\left[ e^{\frac{2i\pi k x}{L_1 + L_2}}a_k + e^{-\frac{2i\pi k x}{L_1 + L_2}}a_k^\dagger\right]
    \label{modeexpansion}
\end{equation}
 leading to the same energy spectrum in equation \ref{diag_eff_H}.

There are a few points to keep in mind while constructing this mode expansion. Firstly, since we are working with finite-length chiral Luttinger liquids, the introduction of the operator $\hat{\chi}$ and its non-trivial commutation with the zero-momentum mode of the Luttinger liquid is essential to ensure the correct bosonic commutators as emphasized in \cite{Geller97}. Secondly, we must take care to quantize the finite-momentum modes of the system over a length $L_1 + L_2.$ Any other quantization results in linear $a_k, a_k^\dagger$ terms in the Hamiltonian, rendering our model nonphysical.

An easier way to go about this problem is to add momentum dependent terms to the mode expansion following \cite{udit&yuval_parafermion}. A general bosonic mode expansion for the chiral fields, is given by
\begin{eqnarray}
\bar{\phi}_{R/L}(x,t) &=& \hat{A}_{R/L}(vt-x) + \hat{\chi}_{R/L} + \nonumber\\
&&\frac{1}{\sqrt{m}}\sum_{n>0}\left(B_{n R/L}\hat{a}_{n}e^{\pm i q_{n}(x \mp vt)} + h.c \right)\nonumber\\
\end{eqnarray}
where $B_{n,R/L}$ are c-numbers, $\hat{A}_{R/L}$ is the zero mode part of bosonic fields, and $\hat{a}_{n}$ is the bosonic annihilation operator in the chiral modes. Imposing Eq.~\ref{bosonicbc} on these chiral bosonic fields gives  $\hat{A}_{R} = \hat{A}_{L} = 2\hat{\eta}/(L_{1} + L_{2})$ and $B_{n,R} = -B_{n,L} = e^{-iq_{n}L_{1}}$, with the chiral fields quantized over $L_{1}+L_{2}$, such that, $q_{n} = 2n\pi/(L_{1}+L_{2})$. $\hat{\chi}_{R/L}$ is the phase operator, with $\hat{\chi}_{R} = -\hat{\chi}_{L} = \hat{\chi}$, such that, $\left[\hat{n}^{SC}_{2},\hat{\chi}\right] = i$. The chiral bosonic modes, $\bar{\phi}_{R/L}(x)$, now are given as
\begin{eqnarray}
\bar{\phi}_{R}(x) &=& \frac{2\hat{\eta} }{L_{1} + L_{2}}(x-L_{1}) +\hat{\chi}
\nonumber \\
~~~&+& \frac{1}{\sqrt{m}}\sum_{k>0}\left( {a}_{k}e^{\frac{2\pi i k}{L_{1} + L_{2}}(x-L_{1})} + {a}_{k}^{\dagger}e^{-\frac{2\pi i k}{L_{1} + L_{2}}(x-L_{1})}\right) \nonumber\\
\bar{\phi}_{L}(x) &=& \frac{2\hat{\eta} }{L_{1} + L_{2}}(x+L_{1}) -\hat{\chi} 
\nonumber \\
~~~&-& \frac{1}{\sqrt{m}}\sum_{k>0}\left( {a}_{k}e^{-\frac{2\pi i k}{L_{1} + L_{2}}(x+L_{1})} + {a}_{k}^{\dagger}e^{\frac{2\pi i k}{L_{1} + L_{2}}(x+L_{1})}\right)\nonumber \\
~~~
\label{Eq:bosonic_field_expansion1}
\end{eqnarray}
such that, equations \ref{comm1} \ref{comm2} are satisfied. Within this approach, the modes are already quantized over length $L_1 + L_2$. Thus, we can directly plug in equation \ref{Eq:bosonic_field_expansion1} into \ref{effectiveHamiltonian1}, obtaining again equation \ref{diag_eff_H}. Note that these approaches are indeed equivalent which is clear from the fact that \(\tilde{\phi}(x) = \bar{\phi}_R(x+L_1) = -\bar{\phi}_L(-x+L_1) \)

\end{widetext}

\bibliography{citations.bib}